\documentclass[twocolumn,superscriptaddress]{revtex4-2}
\usepackage{xcolor}
\usepackage{amsmath, amsfonts, amssymb, graphicx, color}

\graphicspath{{./figures/}}
\renewcommand{\thetable}{\arabic{table}}

\graphicspath{{./Plot/}}

\usepackage{url}
\usepackage{hyperref}

\newcommand{\lastequal}{Corresponding authors. These authors contributed equally.}

\begin{document}

\newcommand{\deftitle}{{Integrating computational detection and experimental validation for rapid GFRAL-specific antibody discovery}}

\title{\deftitle}

\author{Maria Francesca Abbate}
\affiliation{Laboratoire de physique de l'\'Ecole normale sup\'erieure,
  CNRS, PSL University, Sorbonne Universit\'e, and Universit\'e de
  Paris, 75005 Paris, France}
\affiliation{Large Molecule Research, Sanofi, Vitry-sur-Seine, France}
\author{Pierre Toxe}
\affiliation{Large Molecule Research, Sanofi, Vitry-sur-Seine, France}
\author{Nicolas Maestrali}
\affiliation{Large Molecule Research, Sanofi, Vitry-sur-Seine, France}
\author{Marie Gagnaire}
\affiliation{Large Molecule Research, Sanofi, Vitry-sur-Seine, France}
\author{Emmanuelle Vigne}
\affiliation{Large Molecule Research, Sanofi, Vitry-sur-Seine, France}
\author{Melody A. Shahsavarian}
\affiliation{Large Molecule Research, Sanofi, Vitry-sur-Seine, France}
\author{Thierry Mora}
\thanks{\lastequal}
\affiliation{Laboratoire de physique de l'\'Ecole normale sup\'erieure,
  CNRS, PSL University, Sorbonne Universit\'e, and Universit\'e de
  Paris, 75005 Paris, France}
\author{Aleksandra M. Walczak}
\thanks{\lastequal}
\affiliation{Laboratoire de physique de l'\'Ecole normale sup\'erieure,
  CNRS, PSL University, Sorbonne Universit\'e, and Universit\'e de
  Paris, 75005 Paris, France}
\begin{abstract}
The identification and validation of therapeutic antibodies is critical for developing effective treatments for many diseases. We present a computational approach for identifying antibodies targeting GFRAL-specific receptors, receptors implicated in appetite regulation. Using humanized Trianni mice, we conducted a longitudinal study with repeated blood sampling and splenic analysis. We applied the STAR computational method for antibody discovery on bulk antibody repertoire data sampled at key time points. By mapping the output from STAR to single-cell data taken at the last time point, we successfully identified a pool of antibodies, of which 50\% demonstrated binding capabilities. We observed convergent selection, where responding sequences with identical amino acid complementarity determining regions 3 (CDR3) were found in different mice. We provide a catalog of 67 experimentally validated antibodies against GFRAL. The potential of these antibodies as antagonists or agonists against GFRAL suggests therapeutic solutions for conditions like cancer cachexia, anorexia, obesity, and diabetes. This study underscores the utility of integrating computational methods and experimental validation for antibody discovery in therapeutic contexts by reducing time and increasing efficiency.

\end{abstract}

\maketitle

\section*{Significance statement}
Identifying novel therapeutic antibodies can revolutionize the treatment of diseases and help propose new treatments. While in vivo, in vitro and computational methods are currently used, they all suffer from limitations, ranging from human and financial cost, triggering unwanted immune responses and large uncertainty. We combine the strengths of these methods and propose 67 new binders against receptors involved in appetite regulation and stress responses. We design an immunization and analysis pipeline by vaccinating humaninized mice and sampling their repertoires using a combination of bulk and single cell methods. We test and analyze their binding abilities, achieving a 50\% success rate. We show how computational detection combined with experimental validation of antigen-specific antibodies can simplify and speed up antigen discovery.

\section{Introduction}

The discovery of novel therapeutic antibodies has revolutionized the treatment of various diseases
\cite{shawler1985human,schroff1985human,boulianne1984production,carter2006potent}, particularly in the fields of oncology~\cite{henricks2015use,scott2012monoclonal}, immunology~\cite{sevier1981monoclonal},  chronic diseases like migraines~\cite{tso2017anti},
and metabolic disorders~\cite{li2024overview,jamadade2024therapeutic}. 
Antibody-based therapies have enabled targeted treatments that minimize off-target effects compared to traditional drugs \cite{reichert2005monoclonal}. Antibody discovery relies on three main approaches: in vivo, in vitro, and in silico \cite{kennedy2018monoclonal}. In vivo methods, such as hybridoma technology, rely on animal immunization or human patient samples to generate antibodies via the immune system, but they are time consuming and species specific \cite{kohler1975continuous,traggiai2004efficient, wrammert2008rapid}. In vitro methods, like phage display, involve constructing large synthetic antibody libraries from human DNA to identify binding molecules for specific antigens, achieving high affinities in the picomolar range \cite{smith1985filamentous, alfaleh2020phage, reddy2010monoclonal, boder2000directed}. However, in vitro antibodies may have inferior biophysical properties, and non-human antibodies can trigger immune responses \cite{jones1986replacing,shaw1987characterization}. In silico methods use computational tools to design antibodies targeting specific epitopes and optimize their developability by predicting and refining their biophysical properties \cite{sormanni2018third,paul,bennett,dreyer}. They play an increasingly important role in complementing traditional methods and streamlining antibody development. In this study we develop a method that leverages in vivo experiments with computational analyses to select potent antibodies ahead of their testing.

Among the promising therapeutic target for antibody-based therapy is the GDNF family receptor alpha-like (GFRAL) protein, which was identified in 2017 by four independent research groups \cite{mullican2017gfral,yang2017gfral,emmerson2017metabolic,hsu2017non}.
This receptor specifically mediates the effects of the cytokine GDF-15, known for its role in appetite regulation and stress responses \cite{aguilar2021gdf15,macia2012macrophage}. The GDF15-GFRAL pathway is predominantly expressed in the area postrema and nucleus tractus solitarius of the hindbrain. These regions are known for controlling nausea, appetite, and energy homeostasis \cite{tsai2014anorectic}. As GDF15 is released in response to cellular stress or injury, it serves as a ``distress signal'' \cite{johnen2007tumor}, binding to GFRAL to trigger nausea, suppress appetite and reduce food intake \cite{borner2020gdf15}.

Elevated levels of GDF15 have been observed in numerous conditions such as cancer, inflammation, and cardiovascular diseases, correlating with poor prognosis due to severe weight loss and muscle wasting \cite{lerner2016map3k11}. As a consequence, GFRAL has attracted significant attention due to its potential role in treating conditions related to appetite dysregulation, including cancer cachexia \cite{johnen2007tumor,suriben2020antibody,lerner2016map3k11}, anorexia \cite{borner2020gdf15}, obesity \cite{breit2017targeting}, metabolic diseases and diabetes \cite{wang2021gdf15,li2024overview}. Targeting this pathway with antibodies against GFRAL has shown promise in preclinical models in mice of cancer cachexia, leading to restored body weight and improved metabolic health \cite{suriben2020antibody,lee2023gdnf}. Conversely, agonistic GFRAL antibodies may help control obesity by suppressing appetite and reducing food intake \cite{wang2021gdf15,sabatini2021gfral,wang2023gdf15}. This makes GFRAL a key therapeutic target for conditions where appetite regulation and energy balance are disrupted, emphasizing the need for antibodies that specifically bind and inhibit or activate this receptor to mitigate or increase its pathological or beneficial effects.
  
Despite its key role in antibody therapies, few GFRAL-specific antibodies are known.
We aimed to detect GFRAL-specific antibodies through a unique approach combining in vivo experiments, repertoire computational analysis, and in vitro validation. The integration of bulk and single-cell data allowed us to identify and validate antibodies against GFRAL, resulting in a list of 67 binders.
  
\begin{figure*} 
    \centering
    \includegraphics[width=\linewidth]{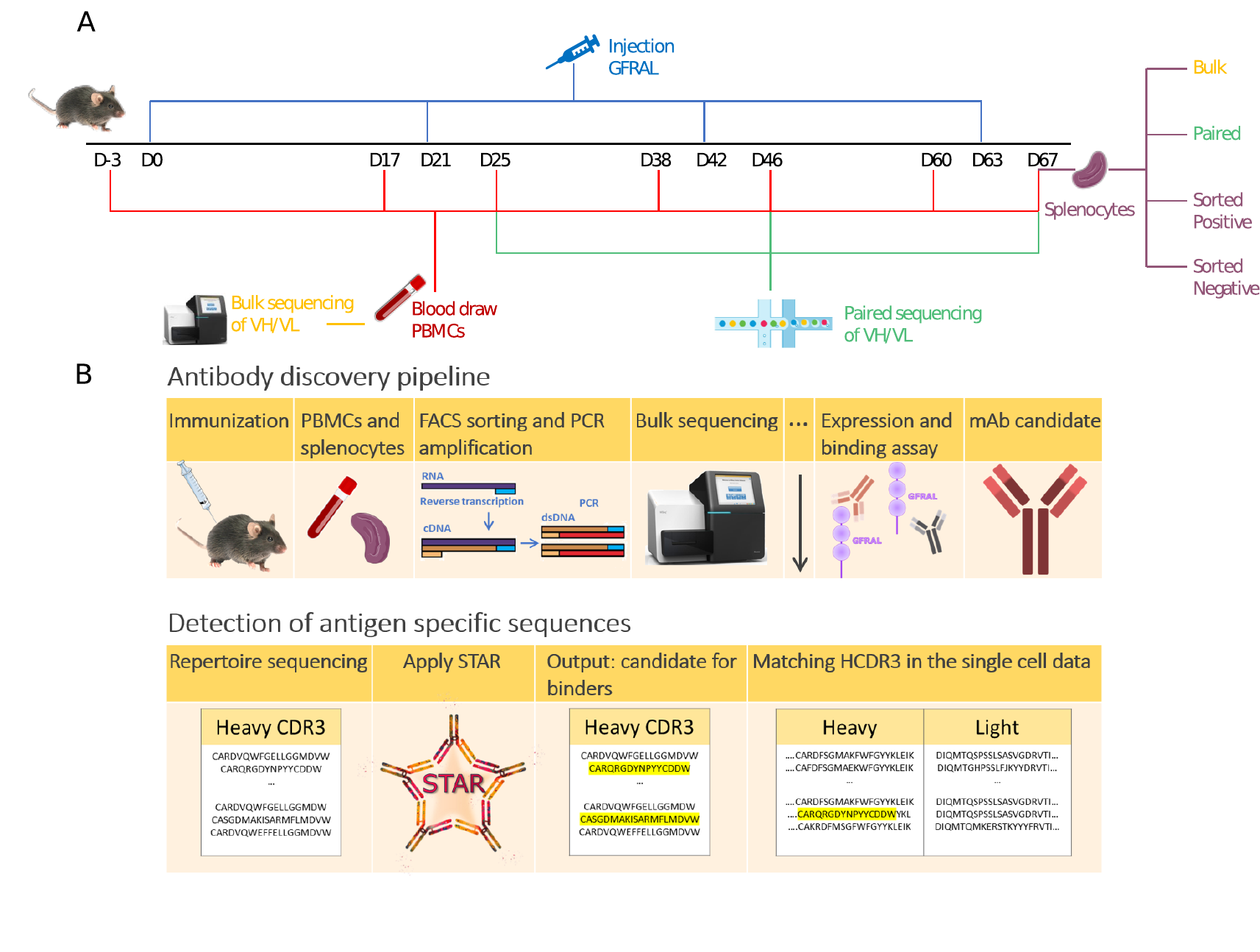}
    \caption{{\bf Experimental setup for the immunization and antibody discovery pipeline.} {\bf A.} Trianni mice were immunized with the GFRAL target at four time points (days 0, 21, 42, and 63). Blood samples were collected on days -3, 17, 25, 38, 46, 60, and 67 to monitor immune responses over time. In addition, spleen samples were obtained from three of the five mice on day 67. Bulk sequencing of both heavy and light chain B cell repertoires was performed for all time points. For select days (25, 46, and 67), a subset of cells were sequenced in single cells to obtain paired heavy and light chains. From the spleen samples on day 67, cells were sorted between GFRAL-positive and GFRAL-negative by flow cytometry, and sequenced in single cells for paired chains. {\bf B.} Overview of the antibody discovery pipeline using STAR to detect the antigen specific sequences. The process begins with immunization with the GFRAL target, followed by isolation of PBMCs and splenocytes and FACS sorting to enrich for B cells. After PCR amplification and sequencing, we applied the computational method STAR~\cite{abbate2024computational} to each sample after the first injection, using only the heavy CDR3 repertoire. The identified sequences were matched to the results of the single-cell sequencing experiments to reveal their corresponding light chain, when found. The selected candidates were expressed and tested for binding using surface plasmon resonance.}
    \label{Cartoon_1_chap5}
  \end{figure*}

\section{Results}


\subsection{Experimental and computational workflow identifies candidate GFRAL-specific antibodies}

Developing human-specific antibodies is essential, as non-human antibodies can elicit immune responses, such as the human anti-mouse antibody (HAMA) reaction, which can diminish therapeutic efficacy and increase safety risks \cite{schroff1985human}. To avoid this issue, we worked with Trianni mice \cite{meininger2016trianni}, which are genetically engineered to produce chimeric antibodies with fully human variable regions (VH,VL).  A group of 5 Trianni mice were immunized with four shots of injection of GFRAL target, taking blood samples four days before and after each injection, and spleen samples from the top three responders after three months. We collected RNA-seq bulk sequences from each time point and paired single cell sequences on day 25, day 46 and day 67 (Fig. \ref{Cartoon_1_chap5}A).  Only samples from the top three responders where analysed. 

 \begin{figure*} 
    \centering
    \includegraphics[width=\linewidth]{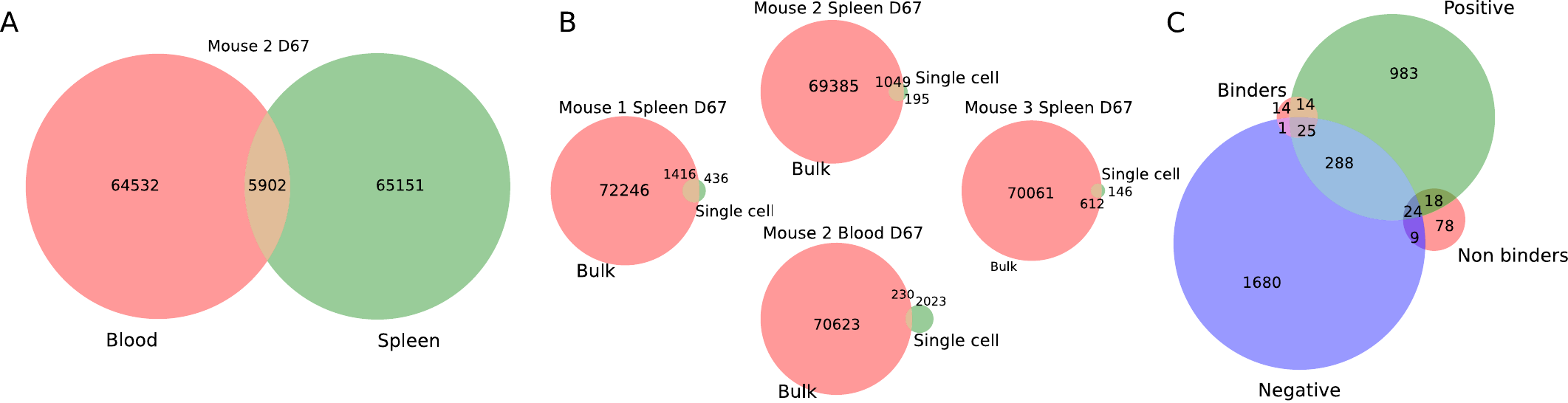}
   \caption{
     {\bf Repertoire overlap.} {\bf A.} Overlap between blood and spleen bulk repertoires for mouse 2 at day $67$.
{\bf B.} Comparison of the overlap between bulk and single-cell sequences from the same tissue and day ($D67$). Specifically, it includes mouse 1 spleen (bulk vs. single cell), mouse 2 (spleen and blood) bulk vs. single cell, and mouse 3 spleen (bulk vs. single cell). This comparison indicates that most information from the single-cell spleen data is captured by the corresponding bulk spleen data, whereas the bulk blood data covers only a limited portion of the single-cell blood repertoire. {\bf C.} Overlap between SPR-validated binding and non-binding sequences versus sequences sorted as antigen-positive and non-positive. Data from all mice. }
    \label{Venn}
  \end{figure*} 

We analyzed the bulk heavy-chain immunoglubulin (IgH) repertoire sequencing data from blood samples of 3 Trianni mice across different time points: days $-3$, $17$, $25$, $38$, $46$, $60$, and $67$, with injection on days $0$, $21$, $42$, and $63$. We observed a total of $3,160,880$ unique nucleotide sequences and $1,530,511$ unique amino acid sequences, combining the data at all timepoints and mice (see SI Table~\ref{tab:SITABLE1}). 
Additionally, to obtain reliable frequency estimates and chain pairing, we incorporated single-cell sequencing data from blood samples on days $25$, $46$, and $67$, with a total of $11,489$ paired sequences, and spleen samples on day $67$, which were sorted into antigen-positive and antigen-negative cells, resulting in a total of $22,898$ successfully paired sequences (Fig. \ref{Cartoon_1_chap5}B).

The overlap between the different datasets (blood vs spleen, single cell vs bulk, antigen-positive vs negative and binders) are summarized in Fig.~\ref{Venn}. Overall, overlap is low because many sequences are rare and seen just once. However, we observed that the single-cell sequences from the spleen were largely included in the corresponding bulk spleen data, whereas the bulk blood repertoire captured only a limited portion of the single-cell blood data (Fig.~\ref{Venn}B). This indicates that, for blood samples, it is beneficial to integrate both bulk and single-cell sequencing to maximize the recovered information. The overlap between surface plasmon resonance (SPR)-validated binding and non-binding sequences, versus those sorted as positive and non-positive (Fig.~\ref{Venn}C), shows a slight enrichment of non-binders among the negative-only fraction.
However, the sorting is not entirely consistent with the SPR assay: some binders are found in the negative sorted set. Also, there is a large overlap between the positive and negative fractions. This is likely due to low B cell receptor expression levels in some cells, which may have led them to be sorted into the negative fraction despite carrying antigen-specific sequences. Since identical sequences are also found in the positive sample, they are likely antigen-specific.

\begin{figure*} 
    \centering
    \includegraphics[scale=0.53]{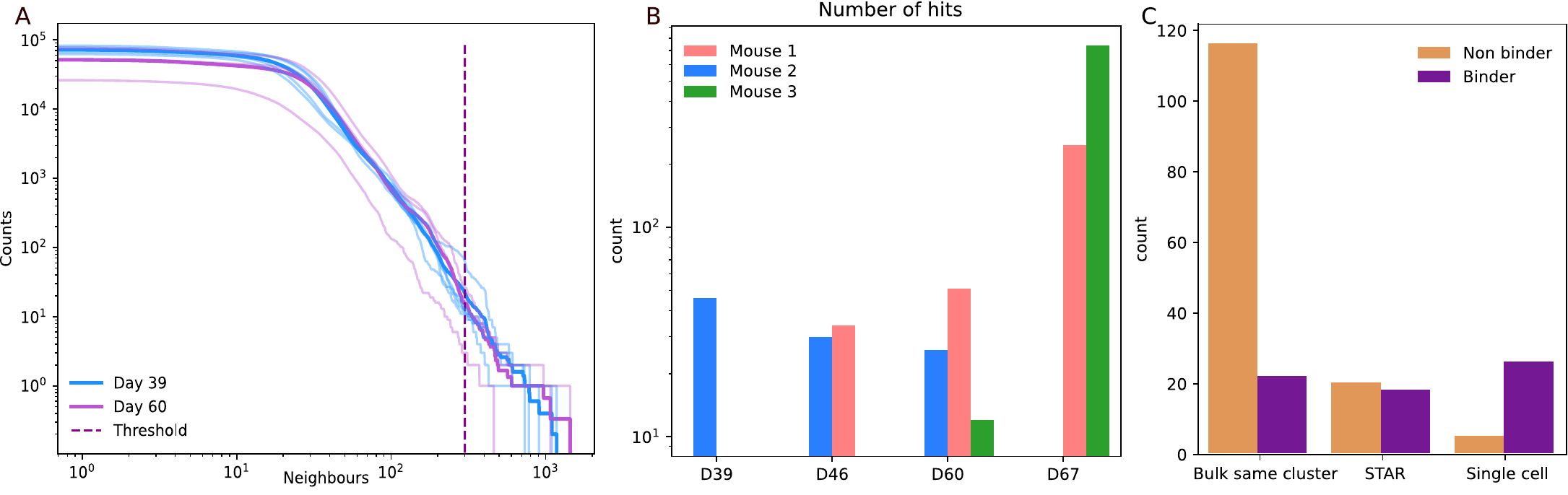}
    \advance\leftskip-1cm
	\advance\rightskip-1cm
    \caption{
        {\bf Number of neighbours, hits and experimental results.} {\bf A.} Distribution of number of neighbours for each mouse at day $39$ and at day $60$. Thick lines show the mean over mice (N=3). According to the STAR method, sequences above the threshold are identified as hits if their cluster contains at least 10 sequences that also exceed the threshold. {\bf B.} Number of hits identified by STAR for each mouse and day from bulk data. {\bf C.} Experimental results by categories: number of experimentally tested sequences identified as binders/non binders by SPR across different categories: the output of STAR, sequences from the bulk repertoire belonging to the same clusters of the sequences as output of STAR and sequences with high frequency in the single cell dataset. Of the $40$ sequences output by the STAR pipeline, $19$ were validated as binders. Additionally, testing of sequences from the single-cell dataset based on frequency yielded $26$ binders, while $22$ binders were identified from bulk repertoire sequences that were absent in the single-cell data, expanding the pool of potential binding antibodies.} 
    \label{Hits}
\end{figure*}

  In the primary workflow, we treat each bulk data time point as independent and analyze them using the STAR method \cite{abbate2024computational}, which looks for signatures of affinity maturation as clumps of closely related sequences in repertoires. Briefly, the method identifies clusters of sequences with statistically overrepresented numbers of neighbours as potential responders. Neighbours are defined as unique CDR3 nucleotide sequences differing by one amino acid. We first cluster heavy-chain CDR3 sequences from bulk repertoire sequencing using this definition of neighbors, and associate each sequence to the cluster it belongs to. For each sequence, we then compute the number of neighboring sequences within the sample (Fig. \ref{Hits}). Sequences that have a sufficient number of neighbors, determined by a threshold set using the STAR method (Fig. \ref{Hits}A), are pre-selected. These sequences are then determined as potential responders (hits), if they belong to a cluster that contains at least 10 such sequences, a parameter that is optimized by STAR. This approach ensures that only clusters with substantial neighbor relationships are considered, helping to exclude isolated clones which, despite having numerous neighbors, do not form dense clusters but rather more linear ones.

Next, we matched these hits with the single-cell sequencing data to identify the corresponding light chains (Fig. \ref{Cartoon_1_chap5}B). For each cluster, we selected a representative sequence with the highest number of neighbors to serve as the reference for matching in the single-cell dataset. We found that $40$ of the sequences identified through STAR were present in the single-cell dataset, and we successfully paired their heavy chains with light chains.

\subsection{Validation of GFRAL specificity by surface plasmon resonance}

We then assayed these sequences for binding using Surface Plasmon Resonance (SPR, see Methods).
Out of $40$ proposed sequences, $19$ were successfully validated as binders by SPR, representing a high success rate (Fig. \ref{Hits}C). Given that these $40$ sequences were drawn from a total population of $1,530,511$ sequences from bulk sequencing, the probability of finding $19$ binders in this group is very low, especially considering that they were required to match sequences from the single-cell dataset.

This two-step approach---combining deep bulk repertoire sequencing with single-cell sequencing---enables us to leverage the depth of bulk sequencing and the chain-pairing precision of single-cell sequencing, thus mitigating the limitations of each method. Single-cell sequencing provides much lower depth than bulk sequencing, which limits its ability to capture rare BCRs and to assess the level of clustering, while bulk sequencing lacks paired chain information.

In addition to the STAR hits, we selected more antibody sequences to be tested by SPR based on two additional criteria: (1) the most frequent sequences from single-cell data, independent of bulk data, which showed an $80\%$ success rate, indicating that frequency after a certain threshold in single-cell data is a good predictor for binders; (2) randomly selected $20$ antibodies from the single-cell dataset, which yielded a much lower binding success rate of $20\%$; and (3) reconstructed antibodies with heavy chains from the same bulk cluster as the initial $40$ hits, but with different mutation levels and frequencies in the heavy CDR3 region, paired with the corresponding light chain from single-cell data. Specifically, for each of the $40$ sequences output of STAR belonging to $40$ different clusters, we took $3$ or $4$ sequences from these clusters, with high, medium and low bulk frequencies and mutation levels compared to the template sequence.  This yielded sequences that were not present in the single-cell dataset, which gave a $19\%$ binding success rate (Fig. \ref{Hits}C). 
Considering that most sequences in this group were absent from the single-cell dataset, this $19\%$ success rate has far greater significance compared to the $20\%$ rate for random single-cell sequences. This finding demonstrates that pairing bulk-derived heavy chains with corresponding light chains from the single-cell data can yield functional, target-specific antibodies not originally found in the single-cell dataset. Consequently, this strategy expands the pool of potential candidates for monoclonal antibody development by generating viable binders beyond those directly observed in the single-cell data.

\begin{figure*} 
    \centering
    \vspace{-1cm}
    \includegraphics[width=\linewidth]{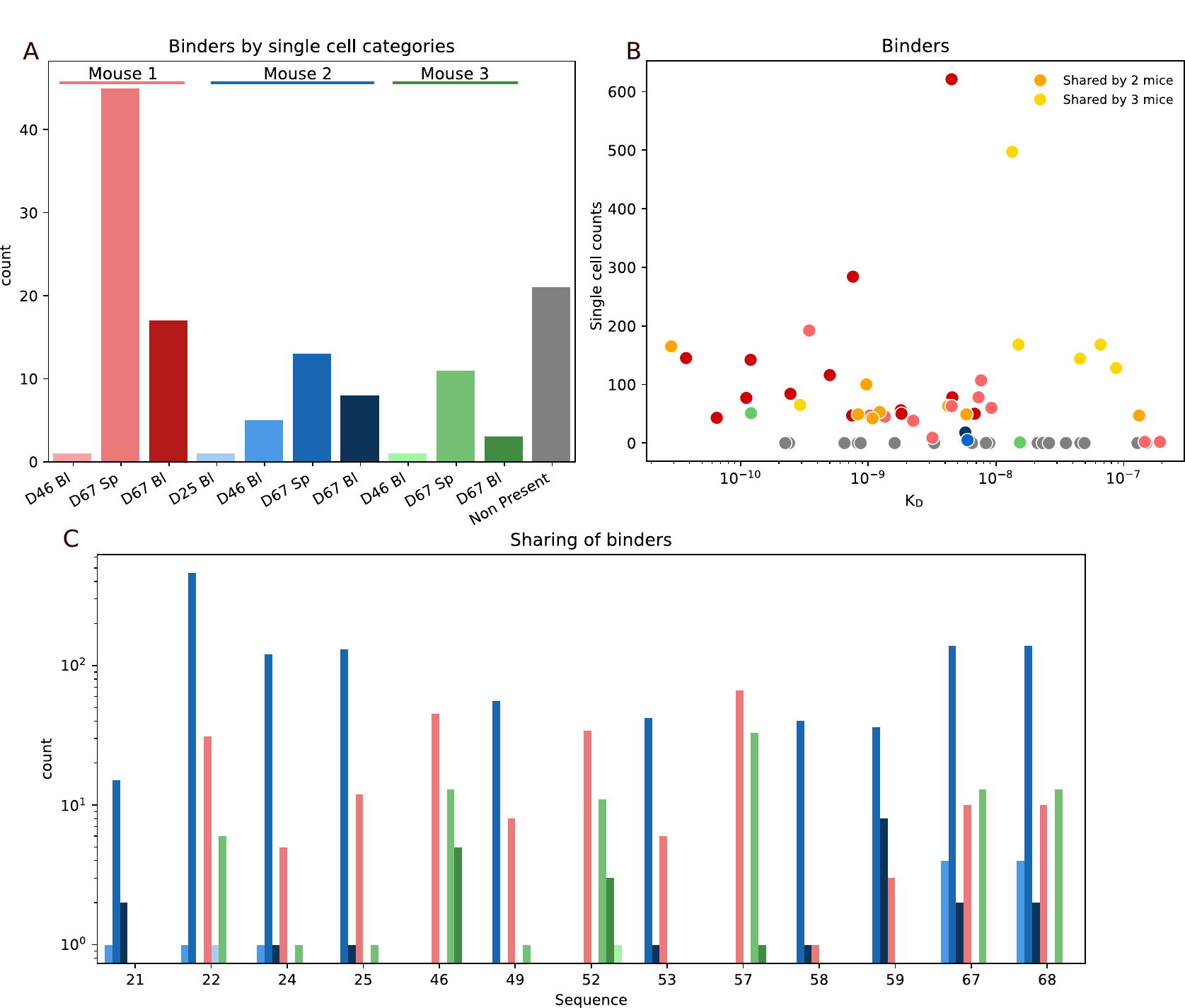}
    \caption{
        {\bf Single-cell analysis of binders by sample type, sequence abundance, and cross-mouse sharing.} {\bf A.} Counts of validated binder sequences in the single-cell dataset, color-coded by tissue source (blood or spleen), mouse, and day. Grey bars indicate synthetic sequences that were not present in the single-cell data. Binder sequences are generally distributed across blood and spleen, with a distinct bias towards spleen-originating sequences in mouse 1.
{\bf B.} Scatter plot comparing the dissociation constant ($K_D$) values of binders with their counts in the single-cell dataset, indicating no correlation between $K_D$ and single-cell abundance. Strong binders (lower $K_D$) may have high or low representation, including those missing from single-cell data (grey points). {\bf C.} Cross-mouse sharing of 13 validated binder sequences, with counts in each sample sources. Shared binders suggest convergent selection across mice.}
    \label{SC_hist}
  \end{figure*}
  
\subsection{Origin and lineage structure of GFRAL antibodies}
  
We investigated the origin, distribution, abundance, and shared characteristics of sequences experimentally validated as binders with SPR.
Fig.~\ref{SC_hist}A shows the sequence counts of the binders color-coded by origin (blood, spleen, mouse, day, and synthetic sequences that were absent from the single-cell dataset, shown in grey). This plot visualizes the primary sources of the binding response, indicating that, apart from mouse 1 where binders predominantly originated from spleen samples, the other mice show relatively balanced contributions from both blood and spleen samples.

Fig.~\ref{SC_hist}B compares each binder’s dissociation constant $K_D$ to its abundance in the single-cell data, revealing only a weak correlation between them (Spearman $\rho=0.26$, $p=0.03$). Strong binders are found across sequences with varying counts in the single-cell data, including those entirely absent from it (grey points), underscoring that affinity is not well predicted by abundance in single-cell data. Fig.~\ref{SC_hist}C shows the abundances of 13 binders that were shared across different mice,  suggesting convergent selection in the immune response.

To understand the patterns of selection during affinity maturation, we reconstructed lineages of antibody sequences from the splenocytes single-cell data sorted into antigen-positive and antigen-negative cells (Fig.~\ref{Lineages}, \ref{fig:lineages_part2}). 
Lineages were inferred using the HILARy software \cite{hilary}, which incorporates heavy and light chain pairing, V and J gene usage, and CDR3 length, and phylogenetic rooted trees were generated with RAxML \cite{raxml}, assigning the inferred germline as the root, and visualized in iTOL \cite{itol}. The lineage reconstruction revealed that, in each mouse, the largest lineages were predominantly composed of antigen-positive sequences (Fig. \ref{Positive_ratio}). This observation is consistent with large lineages originating from the process of affinity maturation, which results in high-affinity binders. To compare lineages across mice, we clustered the CDR3 amino acid sequences of all mice together using single-linkage clustering with links between any pair of sequences differing by at most one amino acid.
While the largest lineages were private to individual mice, meaning that sequences of those lineages didn't cluster with any other from other mice,
we observed medium-sized lineages containing antigen-positive sequences that formed clusters across all three mice (same color in Fig.~\ref{Lineages}, \ref{fig:lineages_part2}). This suggests that different ancestral cells may have independently acquired mutations leading to a shared immune response across mice.

\begin{figure*} 
    \centering
    \advance\leftskip-7cm
	\advance\rightskip-4cm
    \includegraphics[scale=0.72]{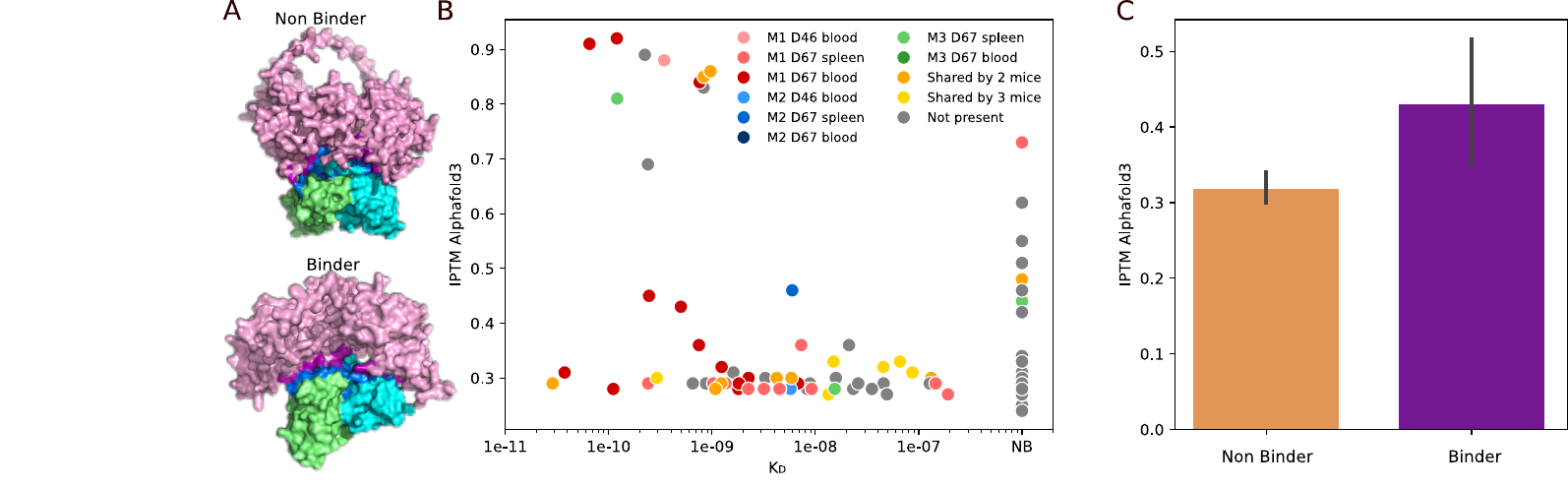}
    \caption{
      {\bf  Predicted binding strength by AlphaFold3.}
      {\bf A.} Examples of predicted structures from AlphaFold3, plotted using PyMol \cite{delano2002pymol}, for three antibody-antigen complexes from each binding category. The antigen is represented in pink, the light chain in green and the heavy chain in cyan. The interface where residues from different proteins are in contact is shown in purple for the antigen (epitope) and in blue for the antibody (paratope). The strong binder shows a more compact interaction with the antigen, while the weak and non-binders exhibit more spread-out configurations. This compact structure in the strong binder is consistent with its higher ipTM score, supporting the ipTM as a proxy for binding affinity.
     {\bf B.} Scatter plot of ipTM scores vs. experimental binding affinity ($K_D$) from SPR. Colors show the origin of sample.
      The plot indicates a moderate negative correlation between ipTM scores for $K_D$ (Pearson $\rho=-0.48$, $p=10^{-5}$), suggesting that higher ipTM values may be associated with stronger binding.
   {\bf C.} Average ipTM scores by binding category confirm that good binders have on average a higher ipTM score.
}
    \label{iptm_hist}
\end{figure*}

\subsection{Alphafold3 analysis of experimentally validated antibody sequences}
To explore the structural basis of antigen-antibody binding, we modelled 137 of the experimentally tested antibodies using Alphafold3~\cite{abramson2024accurate}. Antibodies were categorized based on their experimental binding affinities, determined through surface plasmon resonance (SPR) assays, into  70 binders and 67 non-binders. For each sequence (heavy and light chain) along with its target antigen, we used Alphafold3 to predict the folding and binding structure (Fig. \ref{iptm_hist}A for examples from each binding category) and to produce the interface predicted aligned error (ipTM), a metric that estimates the confidence of the predicted inter-residue interactions at the binding interface. ipTM values range from 0 to 1, where higher values indicate a greater likelihood of a correct structural alignment and potential interaction between the antigen and antibody chains.

The results are summarized in Fig. \ref{iptm_hist}B, which compares the ipTM values between the antibody and the antigen to the measured $K_D$ of all antibodies tested by SPR and shows the average ipTM scores for each binding category. 
Although most sequences show relatively low ipTM scores (suggesting uncertainty or incorrect predictions at the binding interface), binders tend to have higher average ipTM values compared to weak binders and non-binders. This indicates that, while Alphafold3 is not yet fully reliable in modeling the binding interface (average ipTM$\sim 0.3$), and in distinguishing between binders and non-binders, it can still provide valuable structural insights, particularly when the ipTM score is significantly high.

In contrast, Alphafold3 consistently showed high confidence in the relative orientation of the heavy and light chains of the antibody itself, with an average inter-chain ipTM$\sim0.8$ for heavy-light chain interactions. This suggests that while Alphafold can model the overall antibody structure fairly well, it poorly captures its interaction with the antigen. This observation aligns with recent studies~\cite{yin2024evaluation}, highlighting the complexity of modeling antibody-antigen interactions and  the need for enhanced algorithms specifically tuned for inter-protein docking predictions.

\begin{figure*} 
    \centering
    \includegraphics[scale=0.21]{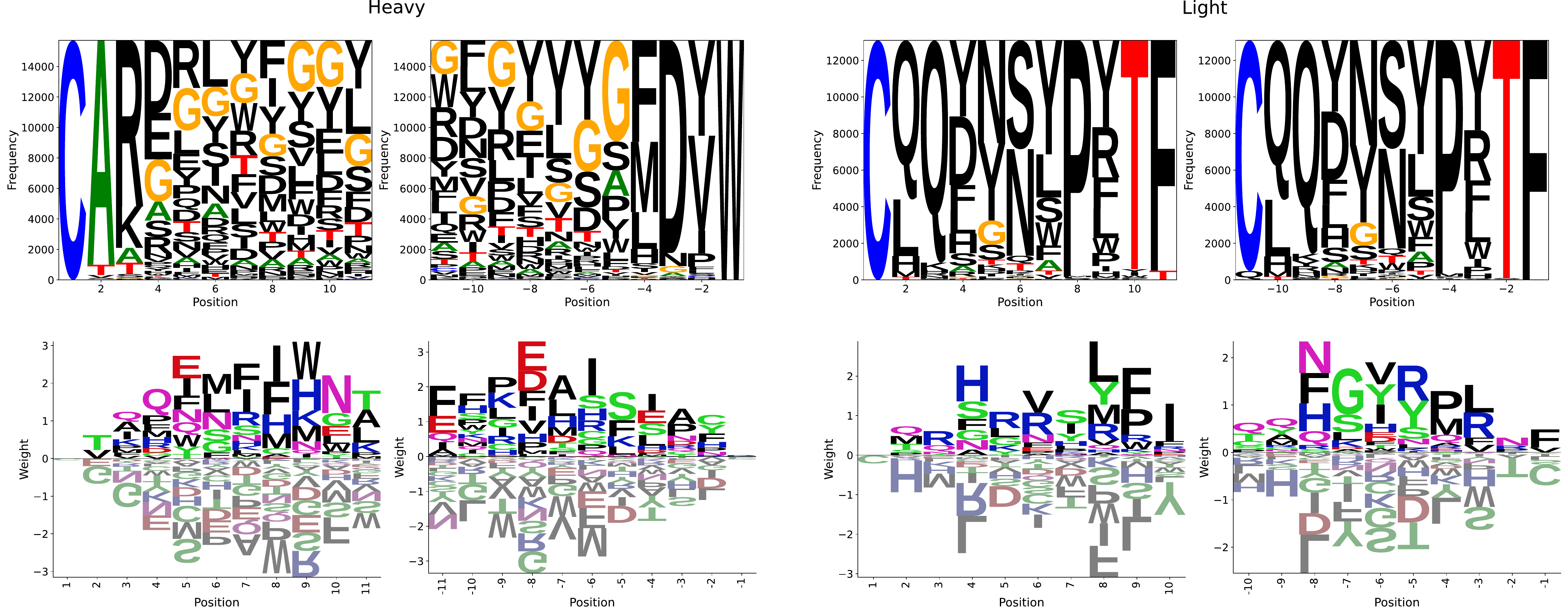}
    \caption{
    {\bf Logistic regression weights and sequence logos for CDR3.} 
    Sequence logos summarizing key features of CDR3 sequences associated with binding. For both heavy (left) and light (right) chains, the top panel shows the logistic regression weights (log odds-ratio) assigned to each amino acid at each position, visualized using Logomaker \cite{tareen2020logomaker}. Letters above the axis correspond to amino acids associated with binding (positive weights), while letters below the axis indicate those associated with non-binding (negative weights). The bottom panels show the sequence logos of all tested binders, providing the positional amino acid enrichment. Together, these representations highlight the specific sequence features predictive of binding in both chains.
    \label{logos}
}

\end{figure*}

\subsection{Sequence features of antigen specificity}
We then asked whether we could extract features of the tested sequences  relevant for binding, focusing on the CDR3 amino-acid sequence of the heavy chain. To determine whether CDR3 sequence information could still be informative about antibody affinity, and used logistic regression to predict its binding status. We used the Left-Right one-hot encoding method \cite{sethna2020population} to transform the sequence into a binary string representing the presence or absence of residues at each CDR3 position, numbered from  the left and from the right of the CDR3 to account for its variable length.
This encoding ensures that positional information is faithfully represented in the model in a symmetric way. The model was trained on 80\% of the antigen positive sorted sequences, tested on the remaining 20\% sequences from the test set, giving an Area Under the Receiving Operating Curve (AUROC) of 0.89 (Fig. \ref{ROC}). While this test suggests that binding is well recapitulated by a simple function of the sequence composition, it is likely confounded by the fact that sequences belonging to the same lineage are both similar in sequence and binding properties, and do not represent independent datapoints that can be separated into training and testing datasets.

To the generalizability of the model more rigorously, we trained it on sequences from two mice and evaluated its performance on a third, independent mouse. This setup avoids potential overlap of clonally related sequences between training and test sets, offering a more stringent evaluation. While reduced, the model still achieved a reasonable accuracy (AUROC=0.73), indicating some conserved features of binding across individuals (Fig. \ref{ROC}).

The logistic regression model assigns weights to each amino acid at different positions across the CDR3, with positive weights indicating amino acids that contribute to binding (binders) and negative weights indicating those that detract from binding (non-binders).
To visualize these contributions, we plotted two sets of sequence logos for the heavy and light chains separately (Fig. \ref{logos}). For each chain, the top panel displays the learned logistic regression weights (log odds-ratio) for each amino acid at each position, with residues promoting binding shown above the axis and those disfavoring binding below. The bottom panel shows the standard sequence logo built from binders only, summarizing the amino acid composition across the CDR3 positions.
These logos reveal the sequence features that are most strongly associated with antigen binding in both heavy and light chains.

\section{Discussion}

In this study, we demonstrated a computational approach for identifying GFRAL-specific antibodies, which we validated experimentally with promising results. 
Our work also provides a list of 67 monoclonal antibodies (mAbs) that were validated to bind to GFRAL, offering a valuable resource for future therapeutic exploration. Given GFRAL’s role as the receptor for GDF-15, a cytokine involved in appetite regulation, the therapeutic potential of these antibodies is significant. GFRAL-targeting antibodies could be developed as antagonists to treat conditions like cancer cachexia and anorexia, where appetite stimulation could reverse severe weight loss. Conversely, agonists could serve as therapeutic tools for obesity and diabetes by reducing appetite. The discovery of these antibodies opens the door for future testing and development of novel drugs that target GFRAL.

However, our current approach has several limitations that warrant future improvements. One major challenge is the need for single-cell data to pair heavy and light chains. While we successfully mapped the heavy chains predicted by our bulk data analysis to the corresponding light chains using single-cell data, this reliance on single-cell information restricts the scalability of the pipeline. An important next step would be to develop a purely computational method to predict these pairings, possibly through machine learning (ML) models that can infer the likely pairing of heavy and light chains based on sequence features. Alternatively, novel bulk sequencing methods that capture both chains such as TIRTL-seq~\cite{pogorelyy2024tirtl} would overcome the scalability issue. 
A given CDR3 heavy chain may be paired with multiple heavy and light chains. Given the depth of single-cell sequencing most of our single-cell sequences revealed a unique pairing but in case of multiple pairing deeper statistical analysis is needed to choose the best candidate. However, given current single cell sequencing costs, the presented matching approach still offers advantages over both bulk and single cell sequencing.

While not a general limitation of the method, our experiments did not use unique molecular identifiers (UMI) in the sequencing protocol \cite{vollmers2013genetic}, which impacts the ability of STAR to reliably call hits. Future experiments should use UMI to achieve the best possible performance.

Another limitation is the complexity of extending the pipeline beyond GFRAL-specific antibodies. While our results are promising, further validation is required to assess the generalizability of the computational pipeline to other targets. Testing this approach on additional targets will help determine how broadly applicable it is in the antibody discovery process. Additionally, improving the statistical models used to match bulk and single-cell data would streamline the process and allow for more accurate predictions without relying heavily on experimental validation at each step.

Despite these limitations, the computational pipeline represents a significant advancement in the efficiency and timeline of the antibody discovery process. By identifying antigen-specific sequences from bulk data, we can narrow down potential binders before moving to more labor-intensive single-cell and experimental validation. This offers considerable time savings and the potential to scale up discovery efforts.

Additionally to identifying responders, we report relatively modest overlap between sequences sampled in the spleen and blood of the same mice. This is likely due to the high diversity of antibody sequences in both datasets. Interestingly, while single cell sequencing overlap with bulk sequencing is relatively larger in the spleen, the larger diversity of the blood repertoire implies a much smaller overlap between the two techniques.

We also explored structural predictions using Alphafold3~\cite{abramson2024accurate}. Overall, while our results demonstrate that high ipTM values can serve as a good indicator of binding when present, the low ipTM values observed for many true binders reflect a broader challenge in using Alphafold3 for precise binding predictions. Given the growing interest in computational methods for antibody discovery\cite{norman2020computational}, these findings suggest that Alphafold3 could be a useful tool for pre-screening, but further refinements are required to improve its predictive accuracy in this context.

In conclusion, this study provides a framework for the computational detection of antigen-specific antibodies, validated through experimental methods. The ability to identify promising binders from bulk repertoire data, combined with insights into convergent selection, offers new opportunities for accelerating the discovery of therapeutic antibodies. With ongoing improvements, this pipeline could play a critical role in advancing the field of antibody therapeutics, not only for GFRAL but for a wide range of clinically relevant targets.

\section{Material and methods}


\subsection{Mice and immunization}

We used five humanized Trianni mice for our experiments. C57BL/6 transgenic humanized Trianni mice were purchased from Taconic Laboratory for experimentation at 6 to 8 weeks of age. During the study, animals were housed 5 per cage, the room was illuminated by lights set to give a 12 h light-dark cycle (on 06.00, off 18.00). The room was air-conditioned by a system designed to maintain an air temperature range of $22 \pm 2C$ and a relative humidity range of $55 \pm 15\%$. Diet and water were given ad libitum. Mouse experiments were carried out in accordance with all federal regulations in an AAALAC-accredited facility. All procedures and protocols were approved by the Sanofi Ethics Committee.
Trianni mice (female, 5 per group) were injected into peritoneal cavity with a 1:1 mixture of $100\mu g$ of GFRAL antigen and Sigma Adjuvant System (S.A.S, Sigma S6322) according to manufacturer’s recommendation. Boost injections were given at later time points (days 21, 42, 63) with the same mixture. All immunization volumes were $200 \mu L$ total.  Blood draws were performed from the submandibular (facial) vein at days -3, 17, 25, 38, 46, 60, and 67 post-initial immunization for further analysis.
Four days after the terminal injection, mice were euthanized. Sera were collected via by intracardiac puncture. Spleens were harvested in Phosphate Buffered Saline 1\% bovine serum albumin for flow cytometry.

\subsection{Bulk and single-cell repertoire sequencing}

Blood samples collected at different time points (day -3, day 17, day 39, day 60 and day 67) were processed for bulk B cell receptor (BCR) sequencing. We performed bulk sequencing on both heavy and light chain, using high-throughput sequencing technologies. Cells were FACS-sorted directly in $700 \mu l$ buffer $RLT + 1\% \beta-OH$  and RNA extraction was performed in bulk using the RNeasy Micro kit $(QIAGEN, \#74034)$. Reverse transcription was made using oligos specifically targeting the constant regions of IgGs (IgG1, IgG2a, IgG2b, IgG2c and IgG3) and IgK mRNA molecules, and cDNAs were amplified by two steps of PCRs with custom oligos. PCR products were sent to a CRO (Eurofins Genomics) for indexing and sequenced using the Illumina NovaSeq $6000 (2x250bp)$ platform. 

Single-cell RNA sequencing (scRNA-seq) was performed using the 10x Genomics system to capture paired heavy and light chain information. Single B cells (Sbc) from blood samples (day 25, day 46 and day 67) as well as Sbc from the spleens were processed following the 10x Genomics protocol for single-cell V(D)J sequencing (Chromium Next GEM Single Cell 5' Reagent Kits v2 Dual Index). Briefly, cells were FACS-sorted in a 96-well plate and centrifuged 5 minutes at 400xg. After removing the supernatant, cells were resuspended with $36.8\mu l$ PBS and mixed with $36.3\mu l$ RT master mix before loading into the chip K for the encapsulation using the chromium controller. Libraries were prepared using the 10x Genomics Single Cell V(D)J Kit, and sequencing was performed on an Illumina NovaSeq 6000 (2x100 bp) platform. This allowed for the generation of paired heavy and light chain data, which were used to match the heavy chains identified from the bulk sequencing. 

The antigen-positive and antigen-negative populations were obtained with the following procedure: after the mouse was sacrificed, splenic B cells were enriched from splenocytes by magnetic negative selection using Pan B Cell Isolation Kit II (Miltenyi; 130-101-638) with addition of anti-IgM Microbeads according to manufacturers’ instructions.  After enrichment, cells were suspended in a buffer solution and stained with 1 $\mu g$ per $10^8$ cells of the following antibodies: Rat anti-mouse IgG1 brilliant violet 421, Rat anti-mouse IgG2a/b brilliant violet 510, rat anti-mouse CD45R brilliant violet 786, rat anti-mouse CD138 and rat anti-mouse IgD, rat anti-mouse IgM, rat anti-mouse CD4, rat anti-mouse CD8 and rat anti-mouse F4/80 brilliant violet 605 (dump channel). For antigen-specific sorting, GFRAL was detected using anti-GFRAL-APC. Cells were separated using BD FACSAria Fusion (BD Biosciences) for flow cytometric cell sorting to isolate IgG antigen-specific B cells or non-specific B cells. Cells were distributed into 10X chips and processed on the 10X instrument.

\subsection{Experimental validation of binders}
To validate the binding capacity of the antibodies identified from the bulk repertoire, we synthesized the corresponding heavy-light chain pairs and performed binding assays.  
Kinetic constants were measured by SPR on a Carterra LSA instrument. An anti human Fc surface was created on a CMDP sensorchip (Carterra) by covalently immobilizing a polyclonal anti huFc antibody (Southern Biotech). Anti GFRAL antibodies were captured at $0.5 \mu g/ml$ for 15 min on anti huFc surface. Serial concentrations of huGFRAL (produced internally) ranging from $0.1$ to $500 nM$ were injected for 5 min. Dissociation was monitored for 30 min and surface was regenerated with a $2 \times 40s$ of $10$ mM glycin-HCl pH$2$. All samples were diluted in running buffer HBS$^-$EP$^+$.
All sensorgrams were double-referenced by subtracting a blank injection and reference surface. Binders were defined based on a $\%$Rmax$>10$. For the binders, curves were fitted with 1:1 model using the Kinetics software (Carterra).

\subsection{Data availability}
The list of validated GFRAL-antibody sequences is available at \href{https://docs.google.com/spreadsheets/d/1kk5N3VKMqvy4FbIhQHC4mAZ_G1zf1_pYoG6-d9GGCGw/edit?usp=sharing}{Antibody SPR Excel file} and the attached {\url{Antibody_SPR.xlsx}} file. The code is available on Github \url{https://github.com/statbiophys/STAR}. The antibody sequence datasets generated in the study are available with restrictions; data can be obtained from the authors upon reasonable request and with permission of Sanofi.

\medskip

\section*{Acknowledgements}
This work was supported by Sanofi, the European Research Council
consolidator grant no 724208 (AMW, TM, MFA), and the Agence Nationale
de la Recherche grant no ANR-19-CE45-0018 “RESP-REP” (AMW, TM,
MFA). We thank Rommel Dadji-Faihun for help conducting SPR experiments,
Anne Brie for preparing the heavy and light chain expression cassettes and 
B\'en\'edicte Onofri for expression and titration of IgG. MFA, MS, MG, PT, NM and EV are or have been Sanofi employees and may hold shares and/or stock options in the company.

\bibliographystyle{unsrt}

\begin{thebibliography}{10}

\bibitem{shawler1985human}
Daniel~L Shawler, RM~Bartholomew, LM~Smith, and RO~Dillman.
\newblock Human immune response to multiple injections of murine monoclonal igg.
\newblock {\em Journal of Immunology (Baltimore, MD.: 1950)}, 135(2):1530--1535, 1985.

\bibitem{schroff1985human}
Robert~W Schroff and Henry~C Stevenson.
\newblock Human immune responses to murine monoclonal antibodies.
\newblock In {\em Monoclonal antibody therapy of human cancer}, pages 121--138. Springer, 1985.

\bibitem{boulianne1984production}
Gabrielle~L Boulianne, Nobumichi Hozumi, and Marc~J Shulman.
\newblock Production of functional chimaeric mouse/human antibody.
\newblock {\em Nature}, 312(5995):643--646, 1984.

\bibitem{carter2006potent}
Paul~J Carter.
\newblock Potent antibody therapeutics by design.
\newblock {\em Nature reviews immunology}, 6(5):343--357, 2006.

\bibitem{henricks2015use}
Linda~M Henricks, Jan~HM Schellens, Alwin~DR Huitema, and Jos~H Beijnen.
\newblock The use of combinations of monoclonal antibodies in clinical oncology.
\newblock {\em Cancer treatment reviews}, 41(10):859--867, 2015.

\bibitem{scott2012monoclonal}
Andrew~M Scott, James~P Allison, and Jedd~D Wolchok.
\newblock Monoclonal antibodies in cancer therapy.
\newblock {\em Cancer immunity}, 12(1), 2012.

\bibitem{sevier1981monoclonal}
E~Dale Sevier, Gary~S David, Joanne Martinis, Walter~J Desmond, Richard~M Bartholomew, and Robert Wang.
\newblock Monoclonal antibodies in clinical immunology.
\newblock {\em Clinical Chemistry}, 27(11):1797--1806, 1981.

\bibitem{tso2017anti}
Amy~R Tso and Peter~J Goadsby.
\newblock Anti-cgrp monoclonal antibodies: the next era of migraine prevention?
\newblock {\em Current treatment options in neurology}, 19:1--11, 2017.

\bibitem{li2024overview}
Jian Li, Xiangjun Hu, Zichuan Xie, Jiajin Li, Chen Huang, and Yan Huang.
\newblock Overview of growth differentiation factor 15 (gdf15) in metabolic diseases.
\newblock {\em Biomedicine \& Pharmacotherapy}, 176:116809, 2024.

\bibitem{jamadade2024therapeutic}
Pratiksha Jamadade, Neh Nupur, Krushna~Ch Maharana, and Sanjiv Singh.
\newblock Therapeutic monoclonal antibodies for metabolic disorders: Major advancements and future perspectives.
\newblock {\em Current Atherosclerosis Reports}, pages 1--23, 2024.

\bibitem{reichert2005monoclonal}
Janice~M Reichert, Clark~J Rosensweig, Laura~B Faden, and Matthew~C Dewitz.
\newblock Monoclonal antibody successes in the clinic.
\newblock {\em Nature biotechnology}, 23(9):1073--1078, 2005.

\bibitem{kennedy2018monoclonal}
Patrick~J Kennedy, Carla Oliveira, Pedro~L Granja, and Bruno Sarmento.
\newblock Monoclonal antibodies: technologies for early discovery and engineering.
\newblock {\em Critical reviews in biotechnology}, 38(3):394--408, 2018.

\bibitem{kohler1975continuous}
Georges K{\"o}hler and Cesar Milstein.
\newblock Continuous cultures of fused cells secreting antibody of predefined specificity.
\newblock {\em nature}, 256(5517):495--497, 1975.

\bibitem{traggiai2004efficient}
Elisabetta Traggiai, Stephan Becker, Kanta Subbarao, Larissa Kolesnikova, Yasushi Uematsu, Maria~Rita Gismondo, Brian~R Murphy, Rino Rappuoli, and Antonio Lanzavecchia.
\newblock An efficient method to make human monoclonal antibodies from memory b cells: potent neutralization of sars coronavirus.
\newblock {\em Nature medicine}, 10(8):871--875, 2004.

\bibitem{wrammert2008rapid}
Jens Wrammert, Kenneth Smith, Joe Miller, William~A Langley, Kenneth Kokko, Christian Larsen, Nai-Ying Zheng, Israel Mays, Lori Garman, Christina Helms, et~al.
\newblock Rapid cloning of high-affinity human monoclonal antibodies against influenza virus.
\newblock {\em Nature}, 453(7195):667--671, 2008.

\bibitem{smith1985filamentous}
George~P Smith.
\newblock Filamentous fusion phage: novel expression vectors that display cloned antigens on the virion surface.
\newblock {\em Science}, 228(4705):1315--1317, 1985.

\bibitem{alfaleh2020phage}
Mohamed~A Alfaleh, Hashem~O Alsaab, Ahmad~Bakur Mahmoud, Almohanad~A Alkayyal, Martina~L Jones, Stephen~M Mahler, and Anwar~M Hashem.
\newblock Phage display derived monoclonal antibodies: from bench to bedside.
\newblock {\em Frontiers in immunology}, 11:1986, 2020.

\bibitem{reddy2010monoclonal}
Sai~T Reddy, Xin Ge, Aleksandr~E Miklos, Randall~A Hughes, Seung~Hyun Kang, Kam~Hon Hoi, Constantine Chrysostomou, Scott~P Hunicke-Smith, Brent~L Iverson, Philip~W Tucker, et~al.
\newblock Monoclonal antibodies isolated without screening by analyzing the variable-gene repertoire of plasma cells.
\newblock {\em Nature biotechnology}, 28(9):965--969, 2010.

\bibitem{boder2000directed}
Eric~T Boder, Katarina~S Midelfort, and K~Dane Wittrup.
\newblock Directed evolution of antibody fragments with monovalent femtomolar antigen-binding affinity.
\newblock {\em Proceedings of the National Academy of Sciences}, 97(20):10701--10705, 2000.

\bibitem{jones1986replacing}
Peter~T Jones, Paul~H Dear, Jefferson Foote, Michael~S Neuberger, and Greg Winter.
\newblock Replacing the complementarity-determining regions in a human antibody with those from a mouse.
\newblock {\em Nature}, 321(6069):522--525, 1986.

\bibitem{shaw1987characterization}
DENISE~R Shaw, MB~Khazaeli, LK~Sun, JOHN Ghrayeb, PETER~E Daddona, S~McKinney, and AF~LoBuglio.
\newblock Characterization of a mouse/human chimeric monoclonal antibody (17-1a) to a colon cancer tumor-associated antigen.
\newblock {\em Journal of immunology (Baltimore, Md.: 1950)}, 138(12):4534--4538, 1987.

\bibitem{sormanni2018third}
Pietro Sormanni, Francesco~A Aprile, and Michele Vendruscolo.
\newblock Third generation antibody discovery methods: in silico rational design.
\newblock {\em Chemical Society Reviews}, 47(24):9137--9157, 2018.

\bibitem{paul}
Paul Pereira, Herv{\'e} Minoux, Aleksandra~M Walczak, and Thierry Mora.
\newblock Energy-based generative models for monoclonal antibodies.
\newblock {\em arXiv preprint arXiv:2411.13390}, 2024.

\bibitem{bennett}
Nathaniel~R Bennett, Joseph~L Watson, Robert~J Ragotte, Andrew~J Borst, D{\'e}Jena{\'e}~L See, Connor Weidle, Riti Biswas, Yutong Yu, Ellen~L Shrock, Russell Ault, et~al.
\newblock Atomically accurate de novo design of antibodies with rfdiffusion.
\newblock {\em bioRxiv}, pages 2024--03, 2025.

\bibitem{dreyer}
Fr{\'e}d{\'e}ric~A Dreyer, Constantin Schneider, Aleksandr Kovaltsuk, Daniel Cutting, Matthew~J Byrne, Daniel~A Nissley, Newton Wahome, Henry Kenlay, Claire Marks, David Errington, et~al.
\newblock Computational design of developable therapeutic antibodies: efficient traversal of binder landscapes and rescue of escape mutations.
\newblock {\em bioRxiv}, pages 2024--10, 2024.

\bibitem{mullican2017gfral}
Shannon~E Mullican, Xiefan Lin-Schmidt, Chen-Ni Chin, Jose~A Chavez, Jennifer~L Furman, Anthony~A Armstrong, Stephen~C Beck, Victoria~J South, Thai~Q Dinh, Tanesha~D Cash-Mason, et~al.
\newblock Gfral is the receptor for gdf15 and the ligand promotes weight loss in mice and nonhuman primates.
\newblock {\em Nature medicine}, 23(10):1150--1157, 2017.

\bibitem{yang2017gfral}
Linda Yang, Chih-Chuan Chang, Zhe Sun, Dennis Madsen, Haisun Zhu, S{\o}ren~B Padkj{\ae}r, Xiaoai Wu, Tao Huang, Karin Hultman, Sarah~J Paulsen, et~al.
\newblock Gfral is the receptor for gdf15 and is required for the anti-obesity effects of the ligand.
\newblock {\em Nature medicine}, 23(10):1158--1166, 2017.

\bibitem{emmerson2017metabolic}
Paul~J Emmerson, Feng Wang, Yong Du, Qian Liu, Richard~T Pickard, Malgorzata~D Gonciarz, Tamer Coskun, Matthew~J Hamang, Dana~K Sindelar, Kimberly~K Ballman, et~al.
\newblock The metabolic effects of gdf15 are mediated by the orphan receptor gfral.
\newblock {\em Nature medicine}, 23(10):1215--1219, 2017.

\bibitem{hsu2017non}
Jer-Yuan Hsu, Suzanne Crawley, Michael Chen, Dina~A Ayupova, Darrin~A Lindhout, Jared Higbee, Alan Kutach, William Joo, Zhengyu Gao, Diana Fu, et~al.
\newblock Non-homeostatic body weight regulation through a brainstem-restricted receptor for gdf15.
\newblock {\em Nature}, 550(7675):255--259, 2017.

\bibitem{aguilar2021gdf15}
David Aguilar-Recarte, Emma Barroso, Anna Guma, Javier Pizarro-Delgado, Luc{\'\i}a Pe{\~n}a, Maria Ruart, Xavier Palomer, Walter Wahli, and Manuel V{\'a}zquez-Carrera.
\newblock Gdf15 mediates the metabolic effects of ppar$\beta$/$\delta$ by activating ampk.
\newblock {\em Cell Reports}, 36(6), 2021.

\bibitem{macia2012macrophage}
Laurence Macia, Vicky Wang-Wei Tsai, Amy~D Nguyen, Heiko Johnen, Tamara Kuffner, Yan-Chuan Shi, Shu Lin, Herbert Herzog, David~A Brown, Samuel~N Breit, et~al.
\newblock Macrophage inhibitory cytokine 1 (mic-1/gdf15) decreases food intake, body weight and improves glucose tolerance in mice on normal \& obesogenic diets.
\newblock {\em PloS one}, 7(4):e34868, 2012.

\bibitem{tsai2014anorectic}
Vicky Wang-Wei Tsai, Rakesh Manandhar, Sebastian~Beck J{\o}rgensen, Ka~Ki~Michelle Lee-Ng, Hong~Ping Zhang, Christopher~Peter Marquis, Lele Jiang, Yasmin Husaini, Shu Lin, Amanda Sainsbury, et~al.
\newblock The anorectic actions of the tgf$\beta$ cytokine mic-1/gdf15 require an intact brainstem area postrema and nucleus of the solitary tract.
\newblock {\em PloS one}, 9(6):e100370, 2014.

\bibitem{johnen2007tumor}
Heiko Johnen, Shu Lin, Tamara Kuffner, David~A Brown, Vicky Wang-Wei Tsai, Asne~R Bauskin, Liyun Wu, Greg Pankhurst, Lele Jiang, Simon Junankar, et~al.
\newblock Tumor-induced anorexia and weight loss are mediated by the tgf-$\beta$ superfamily cytokine mic-1.
\newblock {\em Nature medicine}, 13(11):1333--1340, 2007.

\bibitem{borner2020gdf15}
Tito Borner, Evan~D Shaulson, Misgana~Y Ghidewon, Amanda~B Barnett, Charles~C Horn, Robert~P Doyle, Harvey~J Grill, Matthew~R Hayes, and Bart~C De~Jonghe.
\newblock Gdf15 induces anorexia through nausea and emesis.
\newblock {\em Cell metabolism}, 31(2):351--362, 2020.

\bibitem{lerner2016map3k11}
L~Lerner, J~Tao, Q~Liu, R~Nicoletti, B~Feng, B~Krieger, E~Mazsa, Z~Siddiquee, R~Wang, L~Huang, et~al.
\newblock Map3k11/gdf15 axis is a critical driver of cancer cachexia. j cachexia sarcopenia muscle 7: 467--482, 2016.

\bibitem{suriben2020antibody}
Rowena Suriben, Michael Chen, Jared Higbee, Julie Oeffinger, Richard Ventura, Betty Li, Kalyani Mondal, Zhengyu Gao, Dina Ayupova, Pranali Taskar, et~al.
\newblock Antibody-mediated inhibition of gdf15--gfral activity reverses cancer cachexia in mice.
\newblock {\em Nature medicine}, 26(8):1264--1270, 2020.

\bibitem{breit2017targeting}
Samuel~N Breit, Vicky Wang-Wei Tsai, and David~A Brown.
\newblock Targeting obesity and cachexia: identification of the gfral receptor--mic-1/gdf15 pathway.
\newblock {\em Trends in molecular medicine}, 23(12):1065--1067, 2017.

\bibitem{wang2021gdf15}
Dongdong Wang, Emily~A Day, Logan~K Townsend, Djordje Djordjevic, Sebastian~Beck J{\o}rgensen, and Gregory~R Steinberg.
\newblock Gdf15: emerging biology and therapeutic applications for obesity and cardiometabolic disease.
\newblock {\em Nature Reviews Endocrinology}, 17(10):592--607, 2021.

\bibitem{lee2023gdnf}
Beom~Yong Lee, Jongwon Jeong, Inseong Jung, Hanchae Cho, Dokyung Jung, Jiwon Shin, Jun-kook Park, Eunju Park, Soojeong Noh, Sanghee Shin, et~al.
\newblock Gdnf family receptor alpha-like antagonist antibody alleviates chemotherapy-induced cachexia in melanoma-bearing mice.
\newblock {\em Journal of Cachexia, Sarcopenia and Muscle}, 14(3):1441--1453, 2023.

\bibitem{sabatini2021gfral}
Paul~V Sabatini, Henriette Frikke-Schmidt, Joe Arthurs, Desiree Gordian, Anita Patel, Alan~C Rupp, Jessica~M Adams, Jine Wang, Sebastian Beck~J{\o}rgensen, David~P Olson, et~al.
\newblock Gfral-expressing neurons suppress food intake via aversive pathways.
\newblock {\em Proceedings of the National Academy of Sciences}, 118(8):e2021357118, 2021.

\bibitem{wang2023gdf15}
Dongdong Wang, Logan~K Townsend, Genevi{\`e}ve~J DesOrmeaux, Sara~M Frangos, Battsetseg Batchuluun, Lauralyne Dumont, Rune~Ehrenreich Kuhre, Elham Ahmadi, Sumei Hu, Irena~A Rebalka, et~al.
\newblock Gdf15 promotes weight loss by enhancing energy expenditure in muscle.
\newblock {\em Nature}, 619(7968):143--150, 2023.

\bibitem{abbate2024computational}
Maria~Francesca Abbate, Thomas Dupic, Emmanuelle Vigne, Melody~A Shahsavarian, Aleksandra~M Walczak, and Thierry Mora.
\newblock Computational detection of antigen-specific b cell receptors following immunization.
\newblock {\em Proceedings of the National Academy of Sciences}, 121(35):e2401058121, 2024.

\bibitem{meininger2016trianni}
David Meininger.
\newblock The trianni mouse: The next generation transgenic platform for the isolation of fully human monoclonal antibodies.
\newblock {\em Immunome Research}, 12(S2):38, 2016.

\bibitem{hilary}
Natanael Spisak, Thomas Dupic, Thierry Mora, and Aleksandra~M Walczak.
\newblock Combining mutation and recombination statistics to infer clonal families in antibody repertoires.
\newblock {\em arXiv preprint arXiv:2212.11997}, 2022.

\bibitem{raxml}
Alexandros Stamatakis.
\newblock Raxml version 8: a tool for phylogenetic analysis and post-analysis of large phylogenies.
\newblock {\em Bioinformatics}, 30(9):1312--1313, 2014.

\bibitem{itol}
Ivica Letunic and Peer Bork.
\newblock Interactive tree of life (itol) v6: recent updates to the phylogenetic tree display and annotation tool.
\newblock {\em Nucleic Acids Research}, page gkae268, 2024.

\bibitem{delano2002pymol}
Warren~L DeLano et~al.
\newblock Pymol: An open-source molecular graphics tool.
\newblock {\em CCP4 Newsl. Protein Crystallogr}, 40(1):82--92, 2002.

\bibitem{abramson2024accurate}
Josh Abramson, Jonas Adler, Jack Dunger, Richard Evans, Tim Green, Alexander Pritzel, Olaf Ronneberger, Lindsay Willmore, Andrew~J Ballard, Joshua Bambrick, et~al.
\newblock Accurate structure prediction of biomolecular interactions with alphafold 3.
\newblock {\em Nature}, pages 1--3, 2024.

\bibitem{yin2024evaluation}
Rui Yin and Brian~G Pierce.
\newblock Evaluation of alphafold antibody--antigen modeling with implications for improving predictive accuracy.
\newblock {\em Protein Science}, 33(1):e4865, 2024.

\bibitem{tareen2020logomaker}
Ammar Tareen and Justin~B Kinney.
\newblock Logomaker: beautiful sequence logos in python.
\newblock {\em Bioinformatics}, 36(7):2272--2274, 2020.

\bibitem{sethna2020population}
Zachary Sethna, Giulio Isacchini, Thomas Dupic, Thierry Mora, Aleksandra~M Walczak, and Yuval Elhanati.
\newblock Population variability in the generation and selection of t-cell repertoires.
\newblock {\em PLOS Computational Biology}, 16(12):e1008394, 2020.

\bibitem{pogorelyy2024tirtl}
Mikhail~V Pogorelyy, Allison~M Kirk, Samir Adhikari, Anastasia~A Minervina, Balaji Sundararaman, Kasi Vegesana, David~C Brice, Zachary~B Scott, Paul~G Thomas, SJTRC~Study Team, et~al.
\newblock Tirtl-seq: Deep, quantitative, and affordable paired tcr repertoire sequencing.
\newblock {\em bioRxiv}, 2024.

\bibitem{vollmers2013genetic}
Christopher Vollmers, Rene~V Sit, Joshua~A Weinstein, Cornelia~L Dekker, and Stephen~R Quake.
\newblock Genetic measurement of memory b-cell recall using antibody repertoire sequencing.
\newblock {\em Proceedings of the National Academy of Sciences}, 110(33):13463--13468, 2013.

\bibitem{norman2020computational}
Richard~A Norman, Francesco Ambrosetti, Alexandre~MJJ Bonvin, Lucy~J Colwell, Sebastian Kelm, Sandeep Kumar, and Konrad Krawczyk.
\newblock Computational approaches to therapeutic antibody design: established methods and emerging trends.
\newblock {\em Briefings in bioinformatics}, 21(5):1549--1567, 2020.

\end{thebibliography}

\onecolumngrid
\newpage
\twocolumngrid
\appendix
\renewcommand{\thefigure}{S\arabic{figure}}
\setcounter{figure}{0}
\renewcommand{\thefigure}{S\arabic{figure}}
\renewcommand{\thetable}{S\arabic{table}}

\setcounter{figure}{0}
\setcounter{table}{0}

\begin{figure*}[htbp]
    \centering
    \includegraphics[scale=0.8]{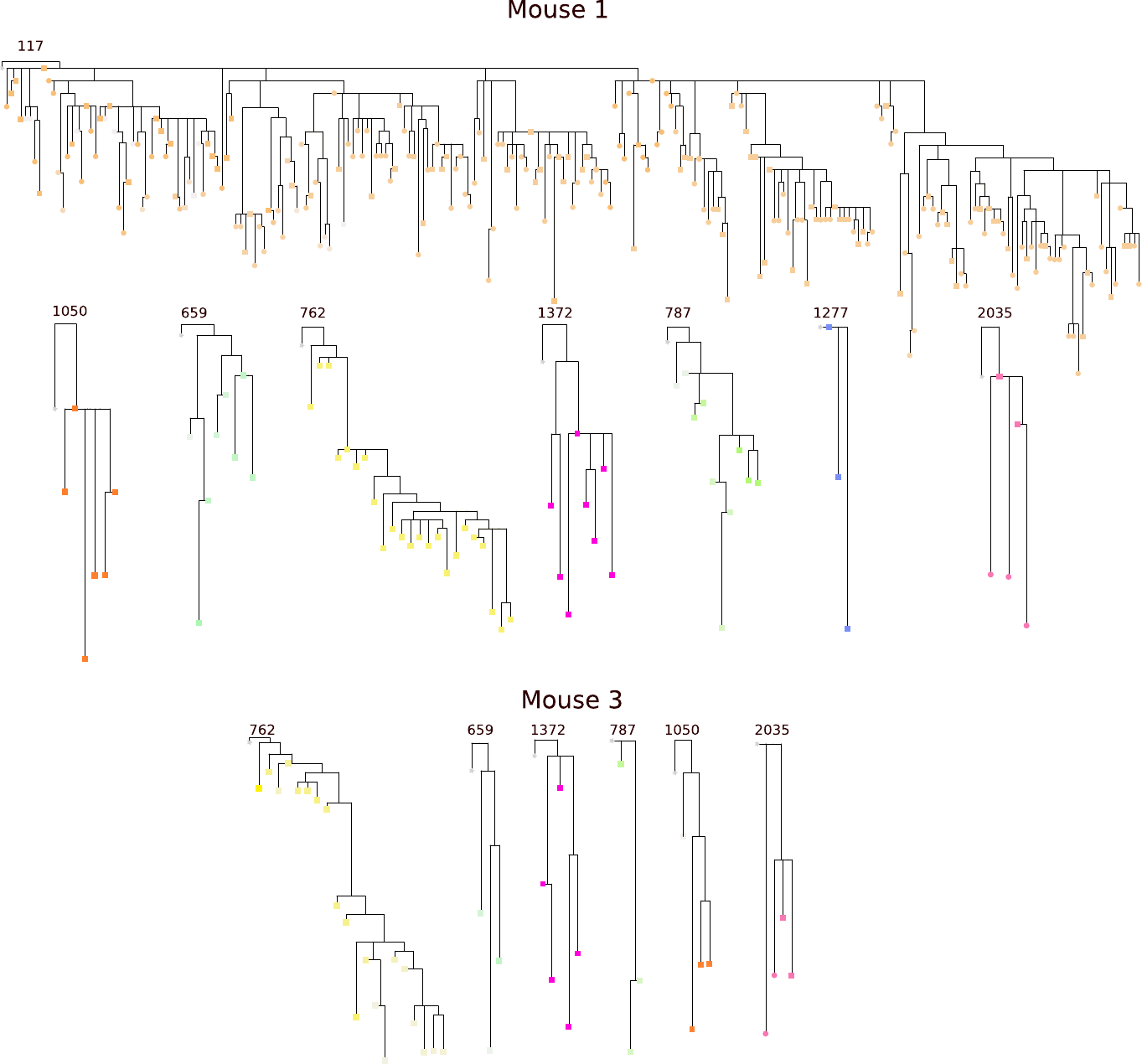}
    \caption{{\bf Lineage trees (Part 1).} Lineage trees of the different mice reconstructed with the sorted positive (squares) and negative (circles) sequences. Each lineage was colored based on the CDR3 amino acid sequence, with shades representing sequence variants. Branch lengths are proportional to the number of mutations. Node sizes represent the count of sequences within the lineage in the splenocytes single-cell sorted data.}
    \label{Lineages}
\end{figure*}
\newpage
\begin{figure*}[htbp]
    \centering
    \includegraphics[scale=0.8]{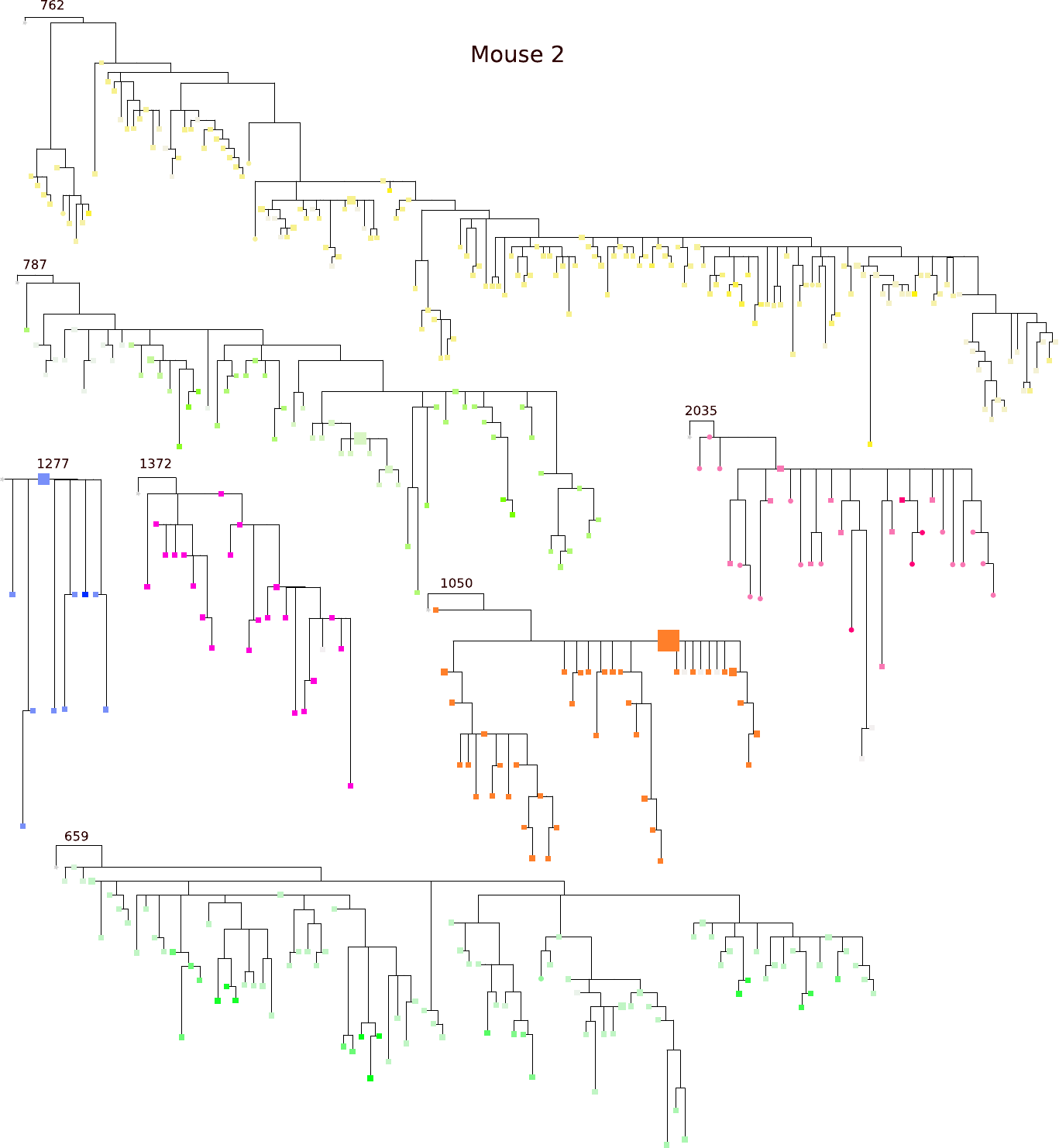}
    \caption{{\bf Lineage trees (Part 2).} Continued from Figure~\ref{Lineages}.}
    \label{fig:lineages_part2}
\end{figure*}

\begin{figure*} 
    \centering
    \includegraphics[scale=0.55]{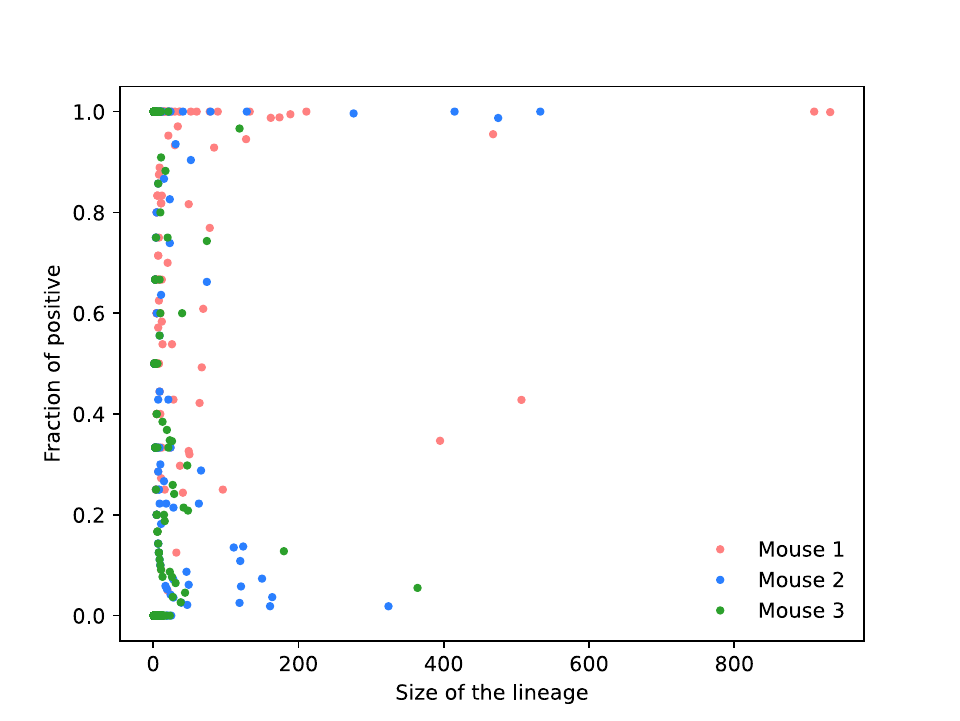}
    \caption{
        {\bf Fraction of positive in the lineage.} 
        Scatter plot showing the relationship between lineage size and the fraction of antigen-specific sequences within each lineage. Each point represents a lineage, with the x-axis indicating the number of sequences in the lineage (on a linear scale) and the y-axis representing the fraction of those sequences that were identified as antigen-specific. The data is color-coded by mouse. 
  A general trend is observed where larger lineages tend to contain a higher fraction of antigen-specific sequences, suggesting a potential enrichment of antigen-specific B cells in expanded clonal families.}
    \label{Positive_ratio}
\end{figure*}

\begin{figure*} 
    \centering
    \includegraphics[scale=0.55]{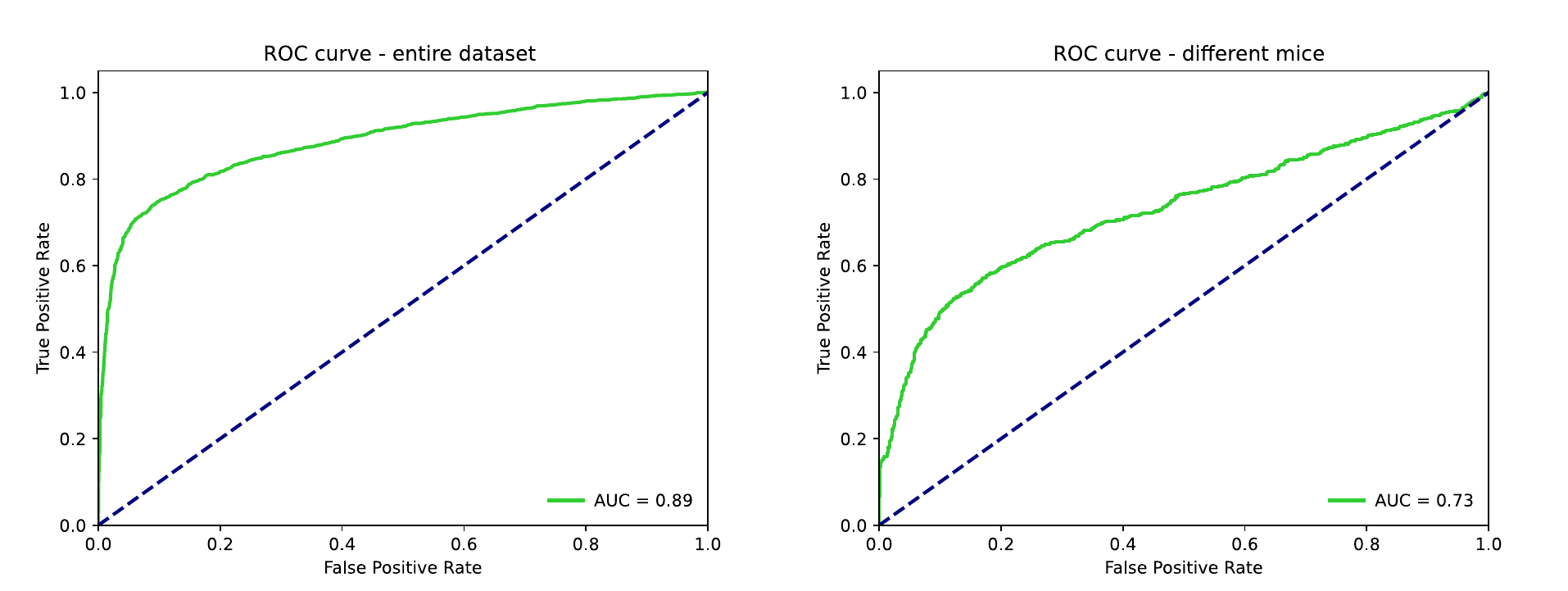}
    \caption{
        {\bf Binding predictions from sequence.} 
       {\bf (A)} Receiver Operating Characteristic (ROC) curve obtained by training a logistic regression model using 80\% of the antigen-specific sorted sequences and testing on the remaining 20\%. Since B cells form clonal lineages, there is a possibility of data leakage between the training and test sets due to closely related sequences being split across them. The model achieves a high performance with an Area Under the Curve (AUC) of 0.89.    {\bf (B)} ROC curve obtained by training the logistic regression model on data from two mice and testing on sequences from a third, independent mouse. This setup eliminates potential data leakage, ensuring a more stringent evaluation of generalization across individuals. The model exhibits reduced performance compared to   {\bf (A)}, with an AUC of 0.73, likely due to inter-mouse variability in the immune repertoire.}
    \label{ROC}
\end{figure*}

\begin{table*}[ht]
\centering
\small
\begin{tabular}{|c|c|c|c|c|c|}
\hline
\textbf{MOUSE} & \textbf{DAY} & \textbf{TYPE} & \textbf{TISSUE} & \textbf{N$^\circ$ of Cell/Reads} & \textbf{N$^\circ$ of Unique CDR3} \\
\hline
Mouse 1 & D17 & Bulk         & Blood  & 3,903,140 & 94,453 \\
Mouse 1 & D25 & Bulk         & Blood  & 3,841,988 & 62,871 \\
Mouse 1 & D25 & Single cell  & Blood  & 1,358     & 1,289  \\
Mouse 1 & D39 & Bulk         & Blood  & 3,972,126 & 83,753 \\
Mouse 1 & D46 & Bulk         & Blood  & 3,878,641 & 68,974 \\
Mouse 1 & D46 & Single cell  & Blood  & 644       & 539    \\
Mouse 1 & D60 & Bulk         & Blood  & 1,276,181 & 26,926 \\
Mouse 1 & D67 & Bulk         & Spleen & 3,675,017 & 73,662 \\
Mouse 1 & D67 & Single cell  & Blood  & 4,272     & 2,604  \\
Mouse 1 & D67 & Single cell  & Spleen & 11,755    & 1,852  \\
\hline
Mouse 2 & D25 & Bulk         & Blood  & 4,161,213 & 68,690 \\
Mouse 2 & D25 & Single cell  & Blood  & 1,358     & 1,289  \\
Mouse 2 & D39 & Bulk         & Blood  & 3,575,899 & 73,227 \\
Mouse 2 & D46 & Bulk         & Blood  & 3,360,544 & 56,469 \\
Mouse 2 & D46 & Single cell  & Blood  & 651       & 565    \\
Mouse 2 & D60 & Bulk         & Blood  & 4,511,141 & 77,013 \\
Mouse 2 & D67 & Bulk         & Blood  & 3,850,030 & 71,053 \\
Mouse 2 & D67 & Bulk         & Spleen & 3,767,658 & 70,434 \\
Mouse 2 & D67 & Single cell  & Blood  & 4,209     & 2,453  \\
Mouse 2 & D67 & Single cell  & Spleen & 7,776     & 1,244  \\
\hline
Mouse 3 & D3  & Bulk         & Blood  & 3,313,716 & 60,136 \\
Mouse 3 & D17 & Bulk         & Blood  & 3,837,689 & 86,623 \\
Mouse 3 & D25 & Bulk         & Blood  & 4,420,976 & 81,634 \\
Mouse 3 & D25 & Single cell  & Blood  & 1,358     & 1,289  \\
Mouse 3 & D39 & Bulk         & Blood  & 3,390,867 & 67,828 \\
Mouse 3 & D46 & Single cell  & Blood  & 727       & 628    \\
Mouse 3 & D60 & Bulk         & Blood  & 3,827,782 & 51,971 \\
Mouse 3 & D67 & Bulk         & Spleen & 3,938,011 & 70,673 \\
Mouse 3 & D67 & Single cell  & Blood  & 1,887     & 1,443  \\
Mouse 3 & D67 & Single cell  & Spleen & 3,783     & 758    \\
\hline
\end{tabular}
\caption{Overview of the number of reads/cells and unique CDR3 sequences for each mouse, tissue, day, and sequencing type.}
\label{tab:SITABLE1}
\end{table*}

\end{document}